\documentclass[preprint,3p]{elsarticle}

\usepackage{graphicx}
\usepackage{multirow} 
\usepackage{amssymb}
\usepackage{amsmath}
\usepackage{enumerate}
\usepackage{subcaption} 
\usepackage{color} 
\usepackage[normalem]{ulem}
\usepackage[utf8]{inputenc}
\newcommand{\degree}{\ensuremath{^\circ}}
\newcommand{\xmax}{\ensuremath{X_{\rm max}\,}}
\newcommand{\Rmax}{\ensuremath{R_{\rm max}\,}}
\newcommand{\xmaxrec}{\ensuremath{X^{\rm Rec}_{\rm max}\,}}
\newcommand{\zhaires}{\mbox{ZHA{\scriptsize{${\textrm{IRE}}$}}{S }}}

\begin{document}

\title{Determination of cosmic-ray primary mass \\
on an event-by-event basis using radio detection}

\author[usp]{Washington R. Carvalho Jr.\corref{cor1}}
\ead{carvajr@gmail.com}

\author[usc]{Jaime Alvarez-Muñiz\corref{cor2}}
\ead{jaime.alvarezmuniz@gmail.com}

\address[usp]{Instituto de F\'\i sica, Universidade de S\~ao Paulo, Rua do Matão  Nr.1371 CEP 05508-090 Cidade Universitária, S\~ao Paulo, Brazil}

\address[usc]{Departamento de F\'\i sica de Part\'\i culas \& Instituto Galego de F\'\i sica de Altas Enerx\'\i as, Univ. de Santiago de Compostela, 15782 Santiago de Compostela, Spain}

\cortext[cor1]{Principal corresponding author}
\cortext[cor2]{Corresponding author}

%%%%%%%%%%%%%%%%%%%%%%%%%%%%%%%%%%%%%%%%%%
\begin{abstract}
We present a new methodology to discriminate between light and heavy ultra-high energy cosmic-ray primaries on an event-by-event basis using information from the radio detection of extensive air showers at MHz frequencies. Similarly to other methods to determine primary cosmic ray composition, the one presented here is based on comparisons between detected radio signals and Monte Carlo simulations for multiple primary cosmic ray compositions. Unlike other methods that first reconstruct the depth of maximum shower development $X_{\rm max}$ to relate it to the nature of the primaries, we instead infer the cosmic-ray composition directly.  The method is most effective in the case of inclined showers that arrive at large zenith angles with respect to the vertical to the ground, where methods based on the determination of \xmax lose accuracy. We show that a discrimination efficiency between 65\% and 80\% can be reached for zenith angles $\theta \gtrsim 60^{\circ}$, even when typical uncertainties in radio detection are taken into account, including shower energy uncertainty. Our methodology could in principle be applied in large and sparse radio arrays, designed with the large radio footprint of inclined showers in mind, to significantly increase the statistics of ultra-high energy cosmic-ray composition studies.

\end{abstract}


\maketitle

%%%%%%%%%%%%%%%%%%%%%%%%%%%%%%%%%%%%%%%%%%
\section{Introduction}

Understanding the nature of Ultra-High Energy Cosmic Rays (UHECR) is crucial to shed light on their origin 
and production mechanisms, and to decipher if the observed suppression of the flux at energies above $\sim 40$\,EeV 
\cite{Augercutoff,Augercutoff-HAS,TAcutoff} is due to propagation effects of the UHECR in the cosmic 
radiation backgrounds \cite{GZK}, 
or to the exhaustion of the sources of UHECR at the highest energies, or possibly to a combination of both effects. 

The state of the art technique for determining both the energy and mass of UHECR is to use observables measured 
with fluorescence detectors (FD) \cite{review_UHECR,Auger_FD}. 
These detect the fluorescence light emitted by the shower as it propagates through the atmosphere 
and reconstruct its longitudinal profile. 
The atmospheric depth of the shower maximum, $\xmax$, is closely related to cosmic ray composition and can be determined with 
an uncertainty of $\sim 20\,{\rm g/cm^{2}}$ \cite{Auger_Xmax,TA_Xmax}. This in turn can be used to infer an average mass 
composition of the cosmic ray flux \cite{Auger_Xmax,TA_Xmax}. 
However, fluorescence detectors can only be used during clear and moonless nights, leading to a small duty cycle of $10\%-15\%$
and to small statistics of the CR flux at the highest energies. 

Detection of the radio emission of extensive air showers was proposed in the 1960's (see \cite{allan} for a review) but was almost completely abandoned shortly after due to technical issues. However, in the last decade there has been a great revival of the radio technique for the detection of UHECR-induced showers in the atmosphere. It is now a well-established air-shower detection technique that is used in several cosmic-ray experiments worldwide, such as the Auger Engineering Radio Array (AERA) at the Pierre Auger Observatory \cite{aera}, 
the Low-Frequency Array (LOFAR) \cite{LOFAR_expt}, TUNKA-REX \cite{tunkarex} and CODALEMA \cite{codalema} among others. Arrays of radio detectors have an almost 100\% duty cycle and for this reason they have been proposed as an alternative to fluorescence telescopes.

The first method to reconstruct $\xmax$ from the information collected with arrays of antennas was developed in the context of the LOFAR experiment \cite{lofarxmax}. It is based on comparisons between the measured electric fields and scintillator data with simulations of proton and iron initiated showers, allowing one to infer the $\xmax$ of the detected event. Variations of this method are currently used by several radio experiments, with a claimed accuracy of $\sim20\,{\rm g/cm^2}$ for showers with zenith angle $\theta \leq 55^\circ$ \cite{lofarxmaxresults-nature,lofarxmaxresults-turin2017,augerradiodetection-turin2017}. These showers have a small radio footprint on the ground that changes rapidly with distance to the shower core. A dense array (distance between antenna elements $D\lesssim500$\,m) is thus required to obtain $\xmax$ with an accuracy comparable to that of FD. This makes the construction of arrays extending over thousands of ${\rm km^2}$, necessary for UHECR detection with high statistics, both challenging and expensive. More inclined showers ($\theta \gtrsim 60^\circ$), however, have a large footprint that can be properly sampled with a more sparse array ($D\gtrsim750$\,m). On the other hand, as we show in Section \ref{sec:xmaxuncertainty} in this paper, the $\xmax$ reconstruction of inclined events using the radio technique has a much larger uncertainty, due to the intrinsic characteristics of inclined showers. This makes inferences of composition inferences with the radio technique that rely on the determination of $\xmax$ very uncertain above $\theta \gtrsim 60^\circ$.

%But, as we will show on section \ref{sec:xmaxuncertainty}, radio $\xmax$ reconstructions of inclined events have a much larger uncertainty. This is due to intrinsic characteristics of inclined showers and makes composition analysis of inclined showers based on this type of $\xmax$ radio reconstruction almost impossible at the highest zenithal angles.

In this work we present the first steps towards a methodology to directly infer the mass composition of UHECR, bypassing the reconstruction of $\xmax$. As previous methods, our approach is also based on comparisons between measured electric fields and those obtained in simulations of showers induced by different primaries. With the methodology presented here we are capable of yielding an efficient composition discrimination of UHECR-induced inclined showers on an event-by-event basis. This conclusion holds even when typical experimental effects, such as radio noise and uncertainties in shower energy and core position are taken into account. As such, this methodology could be used in composition studies of inclined showers, complementing the use of other methodologies %that are more accurate
at smaller zenith angles \cite{lofarxmax}.

This paper is organized as follows: 
Section \ref{sec:radioemission} contains a short review on radio emission from air showers; in Section \ref{sec:zhairessims} we describe in detail the simulations of radio emission from atmospheric showers used in this work, performed with the \zhaires Monte Carlo code. Our approach to discriminate between light and heavy primary UHECR is presented in section \ref{sec:composition}, where we also discuss the impact of experimental uncertainties in the discrimination efficiency of the method. In Section \ref{sec:xmaxuncertainty} we use a variation of the method in \cite{lofarxmax} to reconstruct $\xmax$ from radio observations  to show that its uncertainty increases well above that achieved with FD as the shower zenith angle increases. Outlook and conclusions follow in Section \ref{sec:discussion}.

%%%%%%%%%%%%%%%%%%%%%%%%%%%%%%%%%%%%%%%%%%%%%%%%%%%%%%%%%%%%%%%%%%%%%%%%%%%%%%%%%%%%%%%%%%%%%%%%%%%%%%%%%%%%%%%%%%%%%
\section{Radio emission in atmospheric showers}
\label{sec:radioemission}

Radio emission can be thought of as due to currents induced by the deflection of charged particles in the shower. The induced electric field is approximately proportional to the projection of these currents along a direction perpendicular to the observation direction \cite{toymodel}. The radio emission of extensive air showers is mainly due to the superposition of the geomagnetic \cite{kahnlerchegeo} and Askaryan \cite{Askaryan62} mechanisms. Geomagnetic emission is produced by the deflection of charged particles in the geomagnetic field. Since electrons and positrons are deflected in opposite directions, they both contribute with the same sign to an electric current approximately perpendicular to shower axis, which moves towards the ground along with the shower front. The electric field generated by the geomagnetic mechanism is proportional to the Lorentz force $q\vec{V}\times\vec{B}$, where $q$ is the particle charge, $\vec{B}$ is the geomagnetic field and $\vec{V}$ is the particle speed, which is taken to be approximately parallel to the shower axis. The characteristic polarization of the electric field induced by the geomagnetic mechanism is then approximately parallel to $-\vec V \times \vec B$, and practically independent of observer position. 

The Askaryan mechanism is due to the entrainment of atomic electrons from the medium into the shower flow as the shower evolves, and is mainly due to Compton scattering and knock-on processes such as Moeller and Bhabha scatterings. An excess of electrons over positrons is generated in the shower that is referred to as the charge excess. In this case a current is induced that is approximately parallel to the shower axis and is proportional to the excess of electrons. The electric field generated by the Askaryan mechanism is polarized along the projection of the parallel current onto the plane perpendicular to the observer direction, and is thus approximately zero at the shower core increasing as the observer moves away from it. This leads to an approximately radial polarization towards the shower axis \cite{toymodel,ZHS92}, with a strong dependence on observer position.

While the component of the speed of the charged particles that is parallel to the shower axis is approximately constant and equal to the speed of light $c$, their speed perpendicular to shower axis is much smaller and is mainly due to transverse momenta gained through interactions and the Lorentz force. Although the magnetic force tries to constantly increase the perpendicular momenta of the charged particles, there is a limit to their average perpendicular velocity, called the drift velocity \cite{scholten-driftvelocity}. This limit is roughly inversely proportional to the air density, and it is due to the interactions of the charged particles with the molecules in the medium, which on average tend to randomize the transverse velocity \cite{scholten-driftvelocity}. Since geomagnetic radio emission is due to the current perpendicular to the shower axis, its intensity is approximately inversely proportional to the air density at each stage of shower development. On the other hand, the component of the speed parallel to shower axis is much larger and is unaffected by changes in air density, making the Askaryan contribution to the radio emission practically independent of air density.

The superposition of these two main emission mechanisms, with their different polarizations, makes the pattern (footprint) of the electric field on the shower plane asymmetric with respect to the shower core \cite{zhaires-air}. Since the polarization of the Askaryan component depends on observer position, while the polarization of the geomagnetic component does not, the radio footprint becomes more radially symmetric with an increasing fraction of geomagnetic emission. Also since the geomagnetic component is inversely proportional to air density \cite{scholten-driftvelocity}, and inclined showers develop higher in the atmosphere, the footprint becomes more symmetric %in the plane perpendicular to the shower axis
as the zenith angle of the shower increases \cite{LOFAR_polarization}. 

%%%%%%%%%%%%%%%%%%%%%%%%%%%%%%%%%%%%%%%%%%%%%%%%%%%%%%%%%%%%%%%%%%%%%%%%%%%%%%%%%%%%%%%%%%%%%%%%%%%%%%%%%%%%%%%%%%%%%
\section{Simulations of radio emission}
\label{sec:zhairessims}

In this work we used the \zhaires simulation package \cite{zhaires-air} to calculate the radio emission of UHECR-induced showers in the atmosphere. \zhaires is an AIRES-based \cite{aires} Monte Carlo code that takes into account the full complexity of shower development in the atmosphere, and allows the calculation of the electric field in both the time and frequency domains at different observer positions. \zhaires is based on first principles and does not a priori assume any emission mechanism. However, and as shown in \cite{zhaires-air}, the electric field obtained with \zhaires in the MHz-GHz frequency range is compatible with the superposition of the geomagnetic and charge excess radio emission mechanisms.

In the methodology presented in this work to infer UHECR primary mass, as well as in other methods where $\xmax$ is first reconstructed \cite{lofarxmax}, the position of the shower core can be taken as a free parameter in the minimization process, leading to an optimal core position, for which the simulation best fits the data (see Sections \ref{sec:composition} and \ref{sec:xmaxuncertainty}). On the other hand, varying the core position changes the coordinates of the observers (antennas), where the simulated electric field is compared to the data. A new \zhaires simulation would then be needed each time  to be able to obtain the electric field at each new set of antenna positions (or alternatively a single simulation but with an immense number of antennas). This is not practical from a computational point of view and calls for a fast and accurate method for the calculation of radio emission. For this purpose, in this work we exploit the so-called two-component model addressed in detail in \cite{toymodel}. This model uses as input two \zhaires simulations of a single event: one which includes the geomagnetic field and the other with it artificially turned off. In both simulations the electric field is calculated only in a given number of antennas placed along a line on the ground. This allows one to separate the geomagnetic and Askaryan contributions to the net electric field. These separate contributions, along with their theoretical polarizations, are used to obtain the amplitude and polarization of the peak net electric field at any position on the ground. The two-component model exhibits an accuracy of a few percent when compared to full simulations performed with \zhaires \cite{toymodel}.

We used the two-component model to obtain the electric field in 50 proton and 50 iron-initiated showers at an energy $E=10^{18}$\,eV and zenith angles $\theta =$ 0, 10, 20, 30, 40, 50, 60, 65, 70 and 75$^\circ$. %We also simulated sets of 50 carbon-induced showers with zenithal angles above $\theta=40^\circ$.
For illustration of the method we placed the observers at the site of the LOFAR experiment \cite{LOFAR_expt} (ground altitude of 10 m a.s.l. and a geomagnetic field $|\vec{B}|=49.25\,\mu$T with an inclination of $67.8^\circ$). All showers were injected with an azimuth angle of $\phi=90^\circ$, i.e. arriving at ground from the (magnetic) North. The following parameters were used in the simulations: Thinning level $10^{-5}$, thinning weight factor 0.06, time bin 0.3\,ns and $e^\pm$ (kinetic energy) and $\gamma$ (total energy) cuts of 80 keV. We used SIBYLL 2.1 \cite{SIBYLL} as hadronic model. The electric field needed as input for the two-component model was calculated in $\sim 60$ antennas along a line from the shower core towards the East. The full-band simulations were then filtered between frequencies 30 and 80 MHz and used as input of the superposition model. This bandwidth is commonly used in current radio detection experiments including LOFAR \cite{LOFAR_expt}, AERA\cite{aera} and TUNKA-REX \cite{tunkarex}.

%%%%%%%%%%%%%%%%%%%%%%%%%%%%%%%%%%%%%%%%%%%%%%%%%%%%%%%%%%%%%%%%%%%%%%%%%%%%%%%%%%%%%%%%%%%%%%%%%%%%%%%%%%%%%%%%%%%%%
\section{Inferring primary composition on an event-by-event basis}
\label{sec:composition}

In this section we present a new methodology to infer the primary cosmic-ray composition on an event-by-event basis using information from the radio detection of extensive air showers. Traditionally, $\xmax$ has been used as a surrogate observable for composition \cite{review_UHECR}, and hence reconstructing $\xmax$ is a natural first step in trying to determine cosmic ray composition using the radio technique. On the other hand, radio emission of air showers is a rich and complex phenomena that is very dependent on the geometry and longitudinal profile of the shower and its relationship with the variation of air density and refractive index with altitude. 
These dependencies lead to a strong sensitivity of the pattern of the radio signals on the ground to cosmic ray composition. We argue that this sensitivity, which is also the basis of the $\xmax$ radio reconstruction methods, makes it possible to infer the primary composition of an event even without reconstructing its $\xmax$. 
In the following, we propose to bypass reconstructing the $\xmax$ of the shower and directly infer its primary composition, allowing the method to avoid some of the inherent overlap of the  $\xmax$ distributions of proton and iron-induced showers.

Similarly to the method described in section \ref{sec:xmaxuncertainty} and in \cite{lofarxmax,augerradiodetection-turin2017}, our methodology is also based on comparisons between the electric field measured in several antennas and that predicted in \zhaires Monte Carlo simulations having the same geometry and energy as the detected event, but with different primary compositions. To discriminate the primary composition of a shower event, we firstly perform simulations of 50 proton and 50 iron-initiated showers, 
with random first interaction point in the atmosphere, but with the same energy and geometry (zenith and azimuth angles) of the input event. The measured peak of the radio signal at each antenna $\vec{E}_{\rm data}$, 
defined as the peak of the Hilbert envelope of %each component of
the time-domain signal, is then compared with the peak electric field obtained from simulations $\vec{E}_{\rm MC}$ to calculate $\Delta_s$, defined as the quadratic sum of the differences between measured and predicted electric fields, over all antennas with signal:

\begin{equation}
\Delta^{2}_s=\sum_{i=x,y,z} 
\left(\sum_{\rm antennas}
{\left[ E_{i,{\rm data}} - f_s \cdot E_{i,{\rm MC}} (x-x_{\rm core}, y-y_{\rm core}) \right]^2} \right).
\label{eq:newsigma}
\end{equation}  

Here $f_s$ is an energy scaling factor (see below); $\vec{r}_{\rm core}=(x_{\rm core},\,y_{\rm core})$ is the position of the shower core in the simulation relative to the position of the core of the event to be reconstructed $(x_0,y_0)=(0,0)$ (assumed at the origin of the coordinate system on the ground) and $(x,\,y)$ is the position of each antenna relative to $(x_0,y_0)$. The sums run over the three components of the peak electric field ($E_x, E_y, E_z$) and over all the antennas with signal in the event. Including the polarization in Eq.\,(\ref{eq:newsigma}) could add relevant information for the determination of the primary composition, especially when comparing showers with very different $\xmax$: Showers that develop higher, in a less dense atmosphere, have a higher geomagnetic contribution than those that develop deeper and this changes the ratio between the geomagnetic and Askaryan contributions, changing the observed polarization of the electric field. Although the polarization of the field was explicitly included in $\Delta_s$ to take it into account when discriminating between primary compositions, it is important to stress that the degree to which we can identify heavy from light primaries is not strongly dependent on differences in signal polarization alone. Using Eq.\,(\ref{eq:newsigma}) increases the fraction of correctly discriminated events an average of $\sim 3\%$, if compared to the results using Eq.\,(\ref{eq:sigma}), which does not take into account polarization.

The detected core position and energy of the input event are subject to uncertainties. To account for them, for each of the $s=1,..,N$ simulated proton and iron-induced showers, the position of the core ($\vec{r}_{\rm core}$) and the energy scaling factor ($f_s$) were allowed to vary, leading to different values of $\Delta_s(f_s,\vec{r}_{\rm core})$. The minimum value of $\Delta_s(f_s,\vec{r}_{\rm core})$ for each simulation, denoted simply as $\Delta$, corresponds to the values of $f_s$ and $(x_{\rm core},y_{\rm core})$ for which the simulated shower represents best the measured event. On the other hand, when uncertainties in the core position and energy of the input event are not taken into account, fixed values $f_s=1$ and $\vec{r}_{\rm core}=\vec{0}$ are used in all calculations, leading to a single value $\Delta_s=\Delta$ for each simulated shower. 

Unlike the methods that reconstruct the $\xmax$ of the shower (see Section \ref{sec:xmaxuncertainty}), in this new discrimination method $\xmax$ is not determined. Instead, we compare the distributions of $\Delta^2$ obtained for each simulated composition and infer the most likely composition of the detected event directly. For that purpose, and for the sake of simplicity, we compare the averages\footnote{When obtaining the average values of $\Delta^2$ we do not consider those simulations that have very different footprints from the one in the detected event. This is accomplished by removing the simulations that have $\Delta^2$ above a cut value $\Delta^{2}_{\rm max}= \left< \Delta^{2} \right > - f_{\rm cut} \cdot \sigma_{\Delta^{2}}$. In this work we adopted an optimal value  $f_{\rm cut}=0.15$ in all cases. This cut was implemented because different primary compositions lead to different spreads in the distribution of $\Delta^{2}$. Proton-induced showers have a larger intrinsic fluctuations in their longitudinal profiles if compared to heavier nuclei.  Even if the input event is a proton, there will be a large number of proton simulations that are very different from the input event, increasing the average of the $\Delta^{2}$ distribution for protons, which could create a bias in our method.} $\left<\Delta^{2} \right>_{p}$ and $\left<\Delta^{2} \right>_{Fe}$, which correspond to the average values of $\Delta^2$ obtained using the proton and iron simulations, respectively. The detected event is classified as proton (light) if $\left<\Delta^{2} \right>_p \leq \left<\Delta^{2} \right>_{Fe}$ or iron (heavy) if $\left<\Delta^{2} \right>_p > \left<\Delta^{2} \right>_{Fe}$, i.e. the event is classified as proton or iron depending on what type of primary, on average, induces electric fields that are more similar to the detected fields. Other more sophisticated statistical approaches that benefit from the information available in the distributions of $\Delta^2$ can be applied to classify the events, for instance approaches based on Bayesian statistics or maximum likelihood methods, but they will not be addressed here. Instead, we show below that even with this simple classification criterium we get a large success rate in the determination of the primary composition. 
%%Comment on dependance on hadronic model
It is also worth noting that although we do not reconstruct $\xmax$, our approach should still be dependent on which high-energy hadronic model is used in the simulations. In this work we used SIBYLL 2.1 \cite{SIBYLL}.

\begin{figure}
  \begin{center}
    \includegraphics[width=0.8\textwidth]{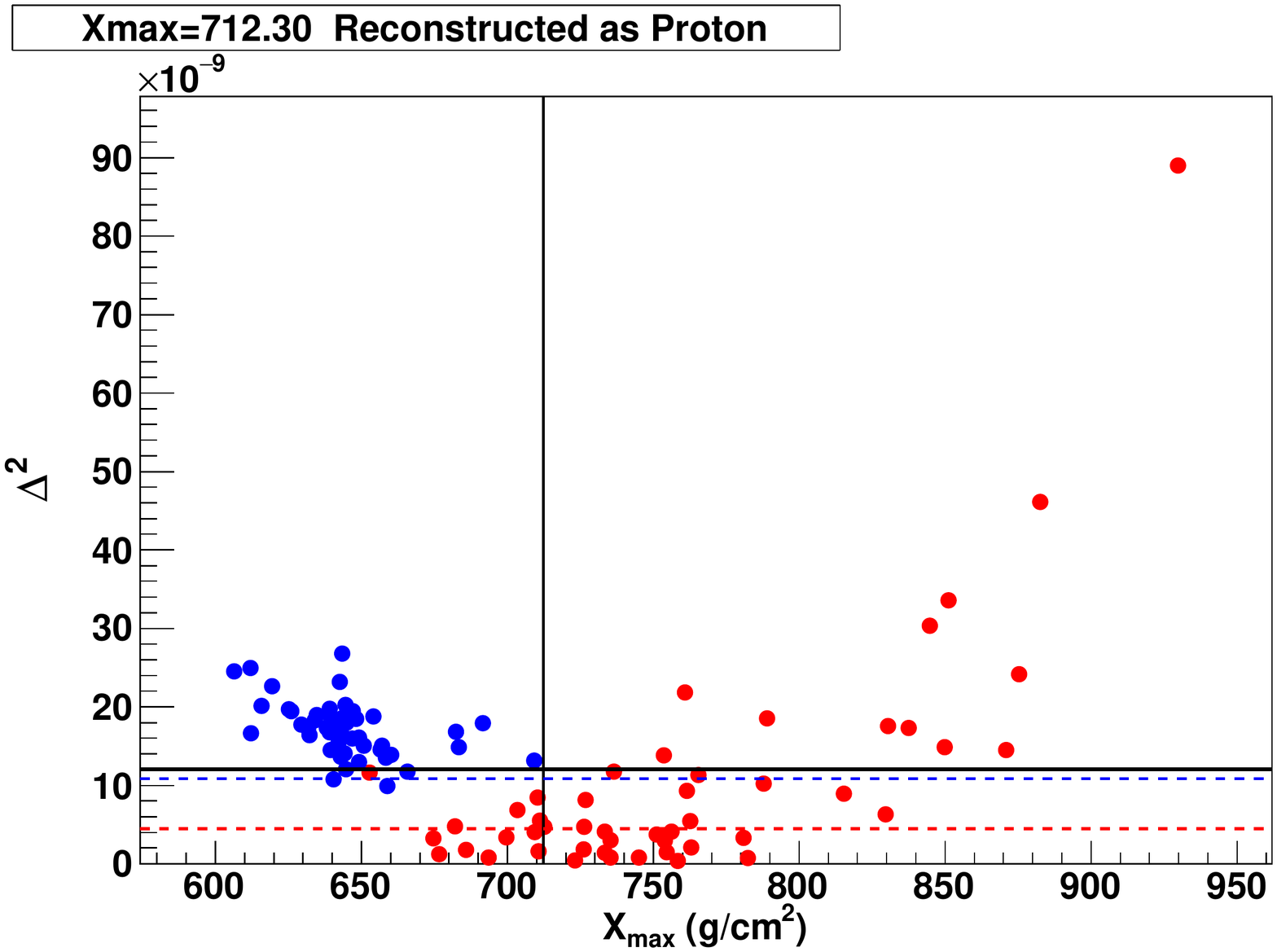}
    \includegraphics[width=0.75\textwidth]{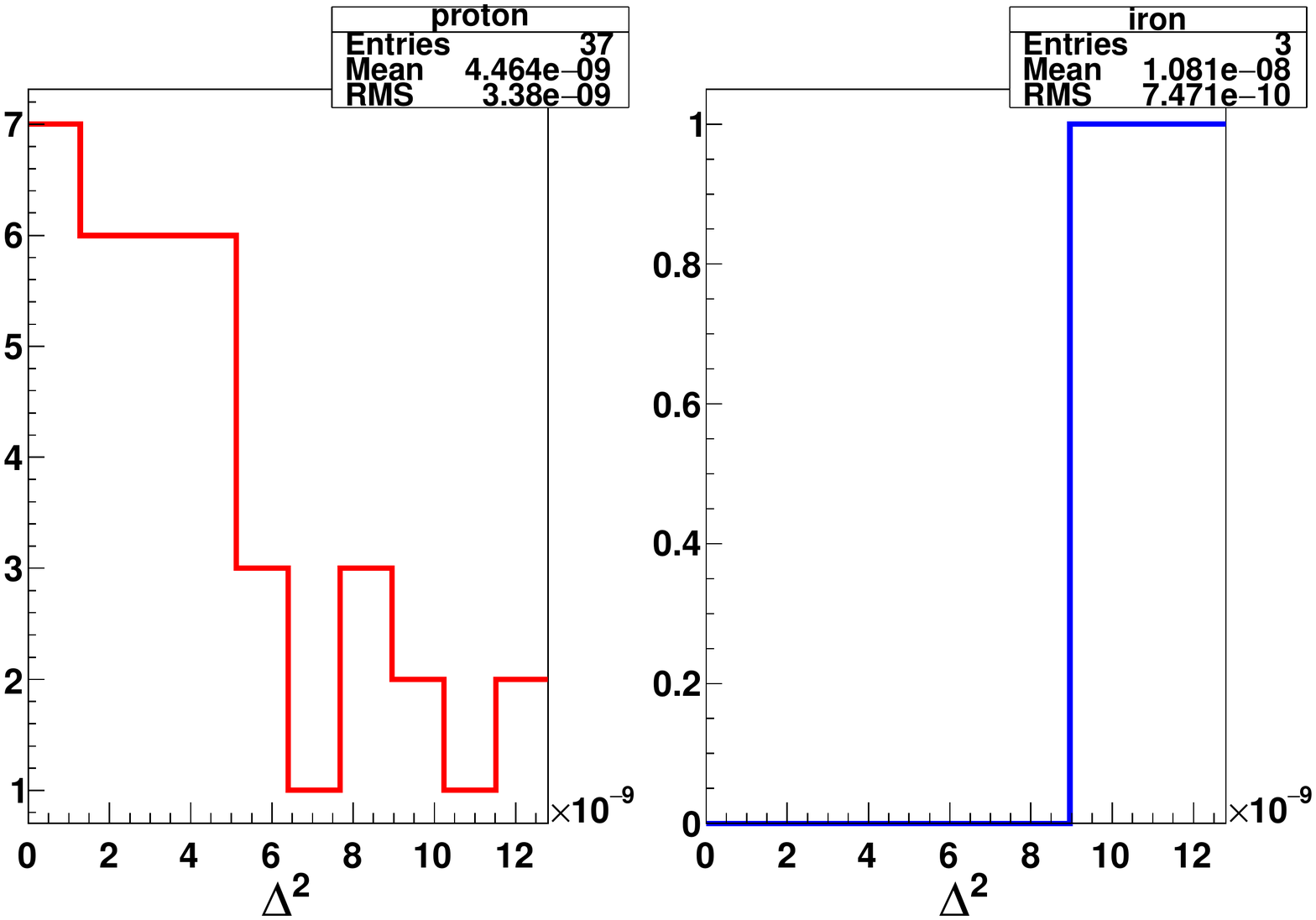}
       \caption{Example of the classification of a (simulated) input proton shower with $E=10^{18}$ eV  and $\theta=65\degree$ using a squared-grid array with distance between antennas $D=500$ m. No detector uncertainties were folded into the simulated input event. Top: $\Delta^{2}$ vs $\xmax$ obtained from each simulation (p in red, Fe in blue). Note that $\xmax$ is neither used nor reconstructed by our method and the top panel serves only to illustrate how $\Delta^{2}$ varies with $\xmax$ and composition. The black solid vertical line represents the $\xmax$ of the input event (in slant g/cm$^2$), while the black solid horizontal line represents the value of the cut in $\Delta^{2}$, above which showers are not considered in the analysis (see text for details). The red and blue dashed lines represent $\left<\Delta^{2} \right>_p$ and $\left<\Delta^{2} \right>_{Fe}$, respectively. Bottom: $\Delta^{2}$ distribution after the cut in $\Delta^{2}$ is applied (left: proton, right: iron). Since $\left<\Delta^{2} \right>_p < \left<\Delta^{2} \right>_{Fe}$ this event was (correctly) classified as proton-like.}
\label{fig:Recexample-nonoise}
\end{center}
\end{figure}

\begin{figure}
\begin{center}

\includegraphics[width=0.9\textwidth]{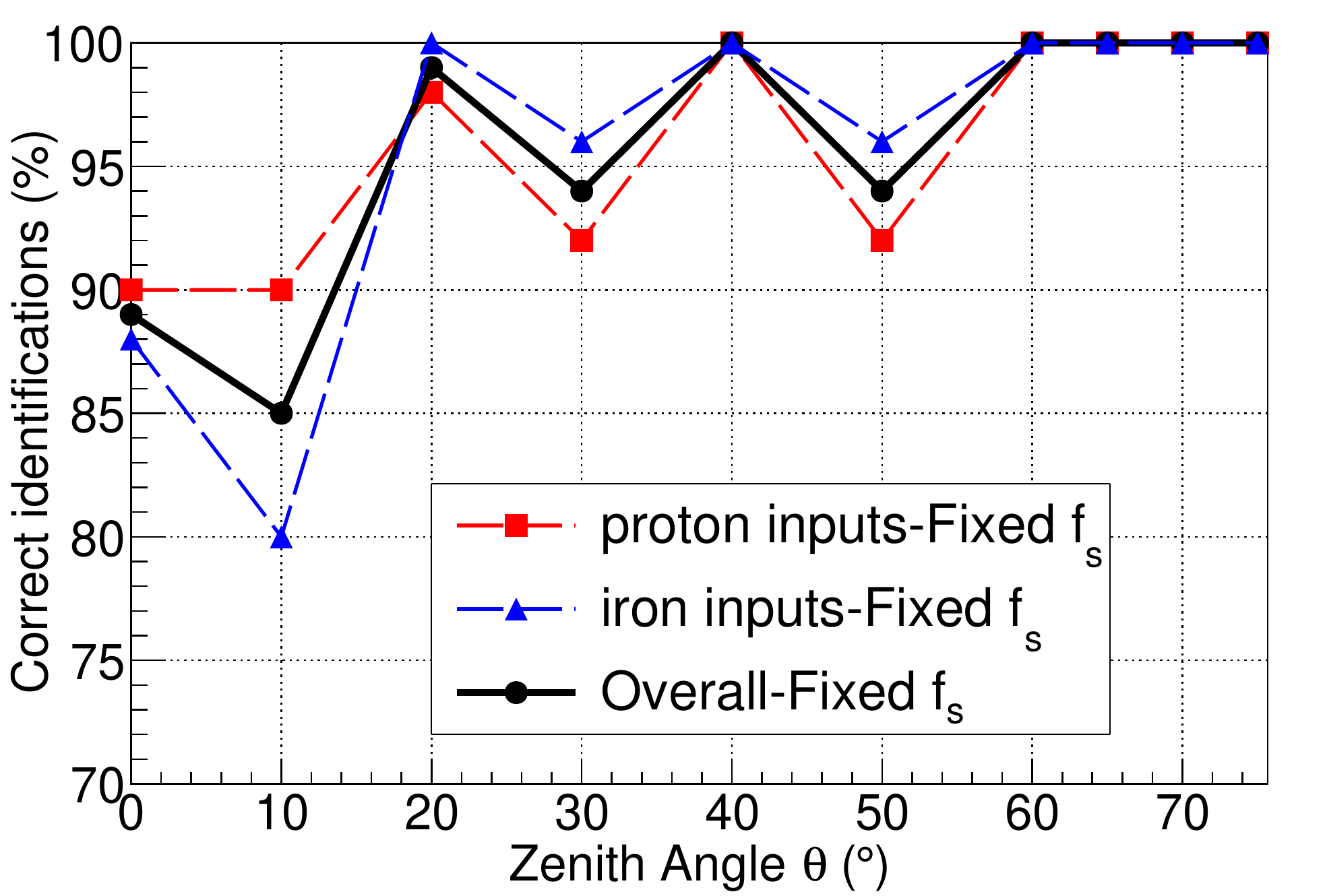}
\caption{Fraction of input events correctly classified as proton or iron as a function of $\theta$ for each simulation set, composed of 50 proton and 50 iron simulated input events per zenith angle. An array with distance between antennas $D=500$ m was used. No detector uncertainties were folded into the input events.}
\label{fig:comperror-nonoise}
\end{center}
\end{figure}

An example of the classification procedure is shown in Fig.\,\ref{fig:Recexample-nonoise}. The input event to be classified is a (simulated) proton shower with $E=10^{18}$ eV and $\theta=65\degree$, triggering a squared-grid array with distance between antennas $D=500$ m. The distributions of $\Delta^2$ obtained when protons or iron are used to infer the composition of the event are shown in the bottom panels in Fig.\,\ref{fig:Recexample-nonoise}. In this case the input event is correctly classified as proton since $\left<\Delta^{2} \right>_p \leq \left<\Delta^{2} \right>_{Fe}$. In fact, when repeating the same procedure for all our 50 proton and 50 iron input events with $\theta=65\degree$, we found that all of them were correctly classified. The fraction of events correctly classified as a function of shower zenith angle $\theta$ is shown in Fig.\,\ref{fig:comperror-nonoise}. One can see that, when no detector uncertainties are taken into account, more than 85\% of the input events have their composition correctly inferred at any zenith angle. When $\theta>60\degree$ this fraction increases to $\sim 100\%$. %$> 95\%$.

%---------------------------------------------------------------------------------------
%---------------------------------------------------------------------------------------

%---------------------------------------------------------------------------------------
%---------------------------------------------------------------------------------------

%%%%%%%%%%%%%%%%%%%%%%%%%%%%%%%%%%%%%%%%%%%%%%%%%%%%%%%%%%%%%%%%%%%%%%%%%%%%%%%%%%%%%%%%%%%%%%%%%%%%%%%%%%%%%%%%%%%%%
\subsection{Detection uncertainties}
\label{sec:uncertainties}
  
In this section we study the effect of detection uncertainties on the ability of the methodology to infer the correct primary composition of events. We took into account the main factors that affect the measurement of radio emission: noise, (galactic) background and uncertainties in the energy of the event and the position of the reconstructed shower core.

\subsubsection{Energy uncertainty}
\label{sec:energy}
 It is well known that iron and proton showers have different missing energies, defined as the energy that goes to high energy muons and neutrinos and that is not deposited in the atmosphere. Because of this, proton induced showers have, on average, more $e^\pm$ ($\sim5\%$ depending on the energy), if compared to iron showers of the same energy, leading to slightly larger electric fields in proton-induced showers. Furthermore, this difference in the deposited energy is much smaller than the characteristic uncertainty in the primary particle energy. To account for this we have used a completely free $f_s$ in the minimization of Eq.~(\ref{eq:newsigma}).

\begin{figure}
\begin{center}

\includegraphics[width=0.9\textwidth]{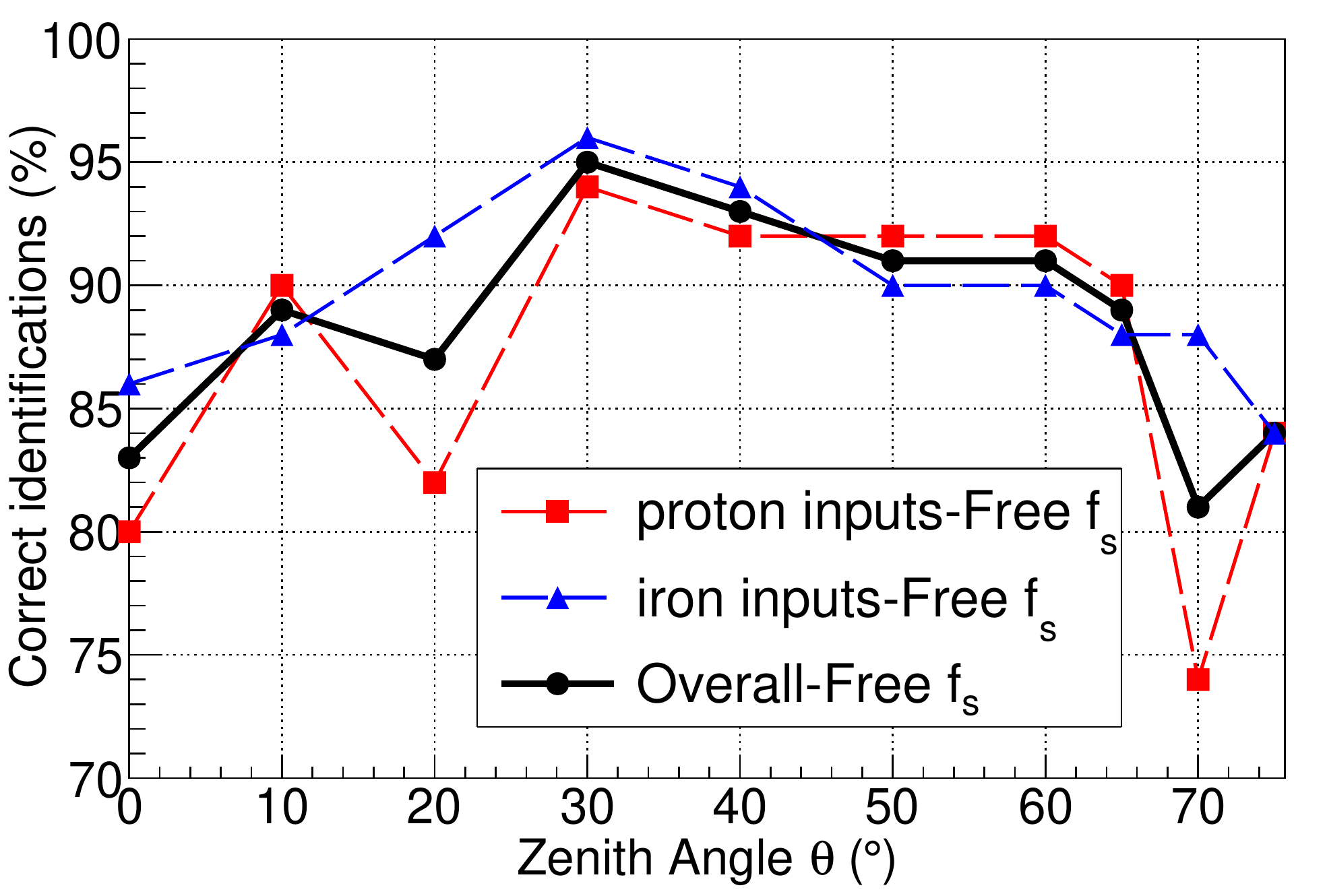}
\caption{Fraction of input events correctly classified as proton or iron as a function of $\theta$ for each simulation set, composed of 50 proton and 50 iron simulated input events per zenith angle. An array with distance between antennas $D=500$ m was used. The only uncertainty that was taken into account was an energy uncertainty adopting a variable $f_s$ parameter in Eq.~(\ref{eq:newsigma}).}
\label{fig:energyalone}
\end{center}
\end{figure}

To isolate the effect of an energy uncertainty on the method, we have included a variable $f_s$ in the reconstructions that use the same ideal detector with distance D=500 m between antennas. The results (including only the uncertainty in energy) can be seen in Fig. \ref{fig:energyalone} that can be directly compared to those shown in Fig. \ref{fig:comperror-nonoise}, where the same detector and other parameters are used but the uncertainty in energy is not accounted for. It is important to notice that allowing a completely free $f_s$ is a worst-case scenario for the method, since this erases not only the average difference between proton and iron showers, but also the shower-to-shower fluctuations of iron and proton showers separately \footnote{In principle one could  consider using an iterative approach, firstly allowing a completely free $f_s$ parameter to obtain two average values of $f_s$ for each shower to be reconstructed: one for comparisons with simulated proton showers, and another for comparisons with iron showers. Then, in a second step, repeat the analysis using only one of these average values of $f_s$, depending on the nature of each simulated shower used for the reconstruction, proton or iron. This would still treat the iron-proton degeneracy due to the missing energy, but would not obliterate the differences due to shower-to-shower fluctuations, which could add relevant information to the discrimination. In any case, we defer such an approach to another work.}.

%---------------------------------------------------------------------------------------
%---------------------------------------------------------------------------------------

\subsubsection{Other detection uncertainties}
\label{sec:otheruncertainties}

To account for the effect of noise on the measured electric field, we estimated a very pessimistic upper limit modeling it as a Gaussian with $\sigma_{\rm noise}=30\,\mu$V/m. A noise amplitude following this distribution is generated for each component of the electric field and for each antenna separately. This simulates the noise temperature contributions from both, the receiver and the sky. The resulting electric field due to noise generated for each antenna is added to the peak electric field of the corresponding antenna of the simulated input event. Also, a fixed electric field background was folded into the input event. For this purpose we used a fixed amplitude of $3\,\mu$V/m and a random isotropic direction for each event, i.e. all antennas detect this same static background component.

We also included the effect of uncertainties in the position of the shower core of the input events by generating a shift $\vec{P}_{\rm error}=(r_{\rm error},\varphi_{\rm error})$ in its core position. For each input event we sample $r_{\rm error}$ from a Gaussian distribution with $\sigma_{r_{\rm error}}=50$ m width \footnote{The uncertainty in core position increases with zenith angle, and can be larger than 50 m for very inclined showers. However we have found that our method is capable of determining the correct core position with a resolution better than 25 m, even at the largest zenith angles. Also, increasing $\sigma_{r_{\rm error}}$ leads to an approximately quadratic increase in computing time. For these reasons, we chose to use a single value of $\sigma_{r_{\rm error}}=50$ m for the whole zenith angle range studied, which is already twice the resolution we have even at $\theta=75^\circ$. We are confident that increasing $\sigma_{r_{\rm error}}$ further for inclined events will not lead to any significant change in our results.} with the angle $\varphi_{\rm error}$ uniformly distributed between 0 and 2$\pi$. We shift the positions of the antenna in the input event by $\vec{P}_{\rm error}$, so that the simulations used for the reconstruction procedure are performed using these dislocated antenna positions. This mimics the effect of applying our methodology to a real event that contains an uncertainty in its measured core position. As discussed in Section \ref{sec:composition}, during the reconstruction procedure we varied\footnote{For this we sweep core positions $(r_{\rm core},\varphi_{\rm core})$ around the origin by varying  $r_{\rm core}$ from 0 to 2.5$\sigma_{r_{\rm error} }$ in steps of $\sim1.5$ m and $\varphi_{\rm core}$ from 0 to 360$\degree$ in 128 steps.} the core position $\vec{r}_{\rm core}$ in Eq.~(\ref{eq:newsigma}) in order to minimize $\Delta^{2}$. In this study we did not take into account shower direction uncertainties, all simulations used for the reconstructions have the same arrival direction as the input event. Also, in order to see the impact of a sparser radio array, we increased the distance between antennas to $D=750$ m in all simulations. Finally, only antennas with a peak amplitude greater than $100\,\mu$V/m were used.

\begin{figure}
\begin{center}
\includegraphics[width=0.9\textwidth]{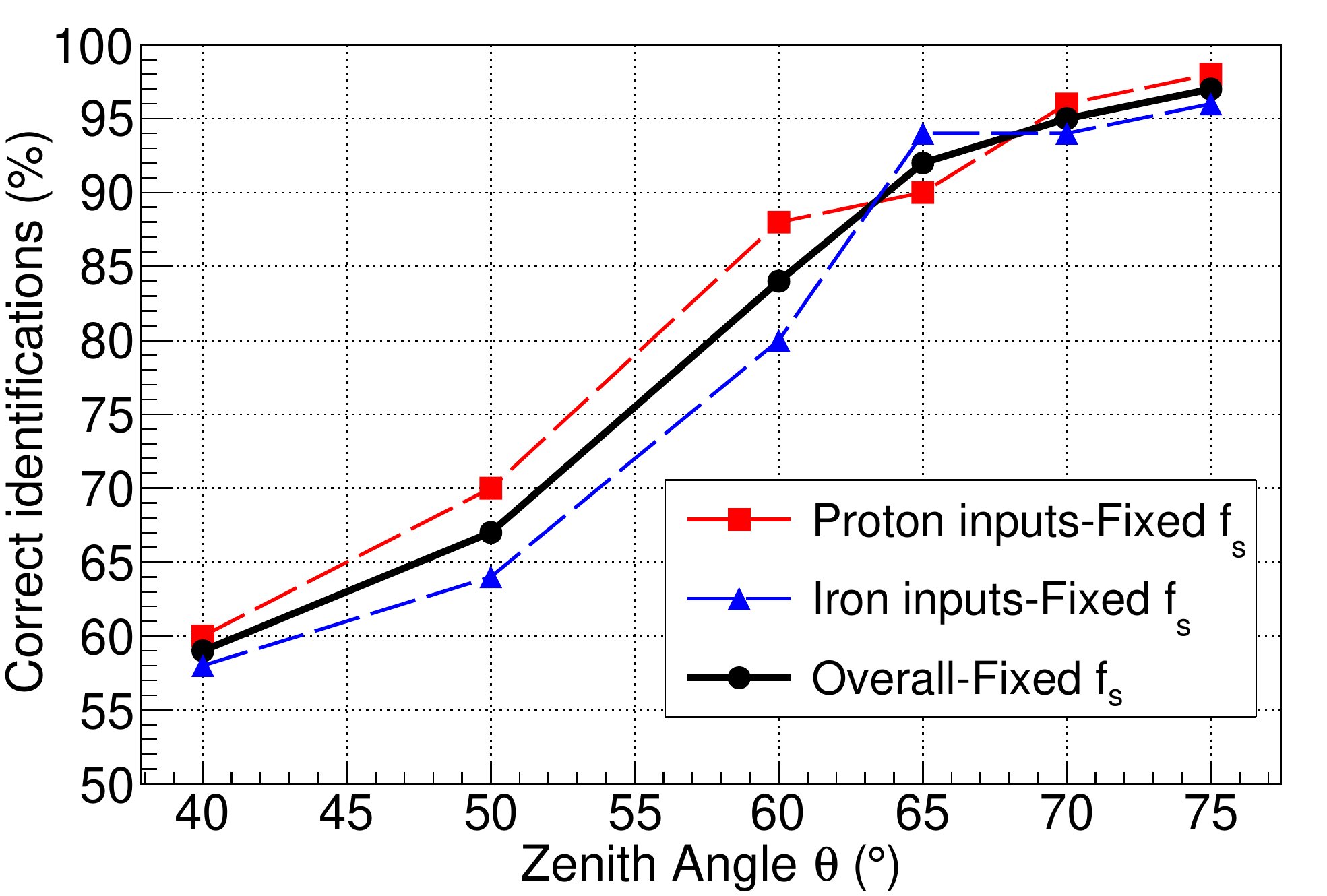}
\caption{Fraction of correctly inferred compositions for each simulation set (each $\theta$). Each set is composed of 50 proton and 50 iron simulated input events. In this case, an array with $D=750$ m was used and all detector uncertainties, except for an energy uncertainty (see text) were folded into the input events. Above $\theta=60\degree$ ($65\degree$), over $\sim 80\%$ ($\sim90\%$) of the events had their composition correctly discriminated by the method.}
\label{fig:comperror}
\end{center}
\end{figure}

 In order to isolate the effect of these other uncertainties on the method, we included them in the simulations, but at first disregarding any energy uncertainty by using a fixed value $f_s=1$ in Eq.\,(\ref{eq:newsigma}). In Fig.\,\ref{fig:comperror} we show the fraction of events that had their primary composition correctly inferred by the method as a function of zenith angle, when all detection uncertainties except for the energy uncertainty were included. We can see that the method becomes more efficient at higher zenith angles, reaching a 90\% correct discrimination fraction above $\theta=65\degree$. 

%---------------------------------------------------------------------------------------
%---------------------------------------------------------------------------------------

\begin{figure}
\begin{center}
\includegraphics[width=0.9\textwidth]{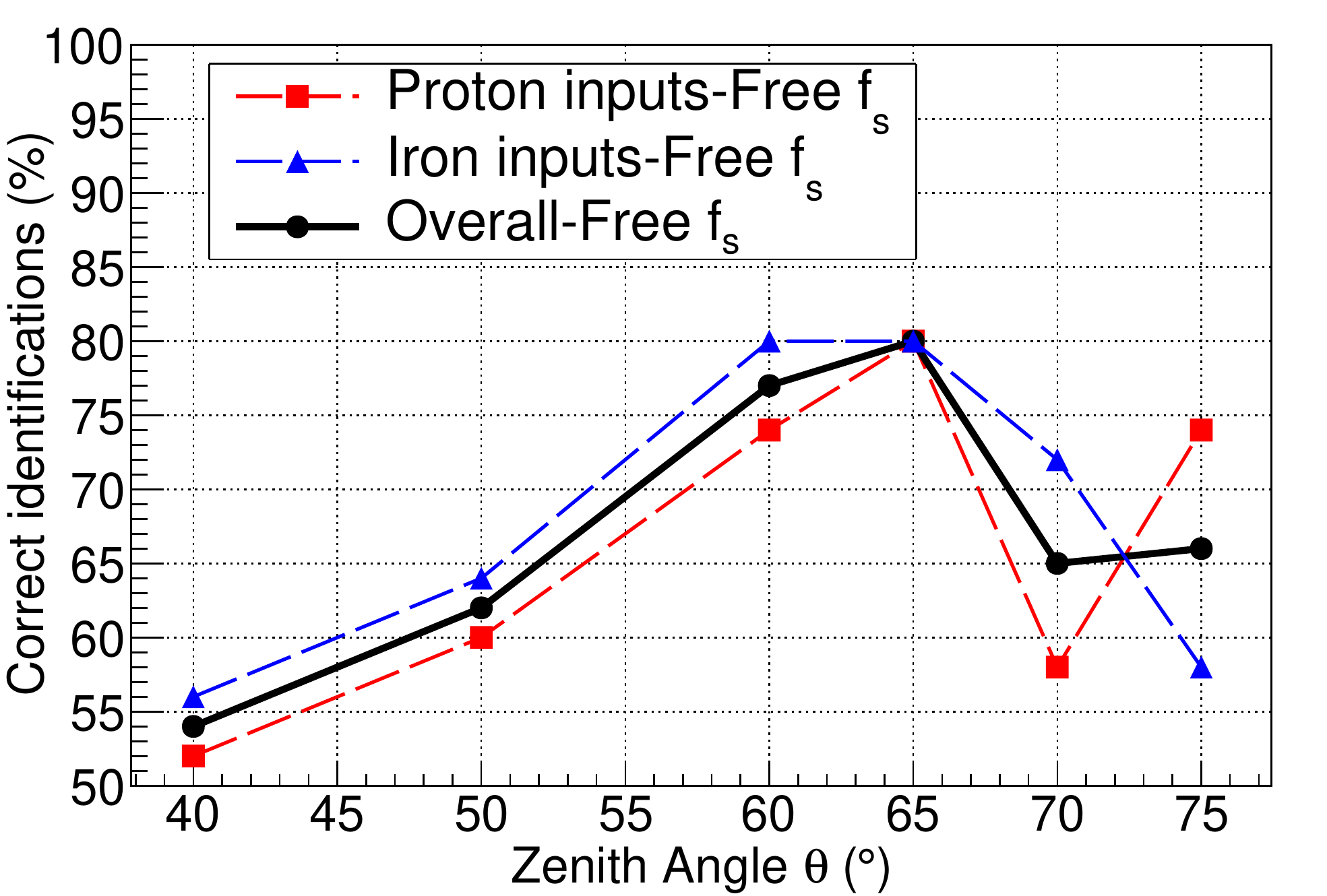}
\caption{Fraction of correctly inferred compositions for each simulation set (each $\theta$). Each set is composed of 50 proton and 50 iron simulated input events. In this case, an array with $D=750$ m was used and all detector uncertainties, including an energy uncertainty (see text) were folded into the input events.}
\label{fig:comperror-all}
\end{center}
\end{figure}

We then included the energy uncertainty along with all the other uncertainties mentioned above. Our results for the  fraction of events that had their composition correctly inferred can be seen on Fig.\,\ref{fig:comperror-all}. Here all uncertainties have been included, i.e. we included a completely free $f_s$ parameter in Eq.\,(\ref{eq:newsigma}). One can see that by including the energy uncertainty a best efficiency of $\sim80\%$ is reached at $\theta=65\degree$, decreasing to $\sim65\%$ at $\theta=70^\circ$ and $75\degree$. Nevertheless, our method should still be the better option for obtaining the primary composition of events above $60\degree$, if compared to $\xmax$ radio reconstructions, due to the large uncertainties and systematic errors when reconstructing the $\xmax$ of inclined events using the radio technique, as will be discussed in Section \ref{sec:xmaxuncertainty}.

\begin{figure}
  \begin{center}
    \includegraphics[width=0.9\textwidth]{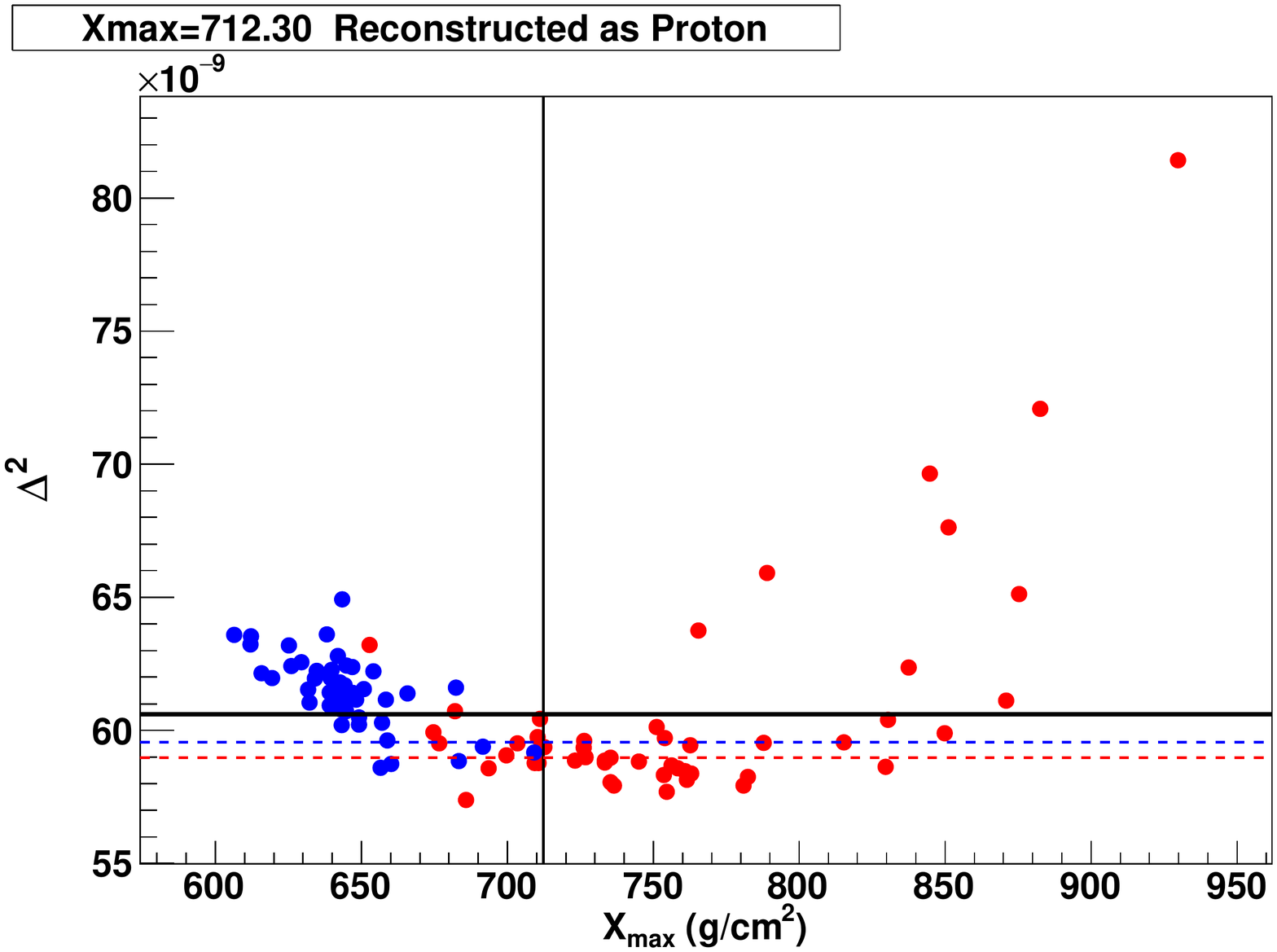}
    \includegraphics[width=0.9\textwidth]{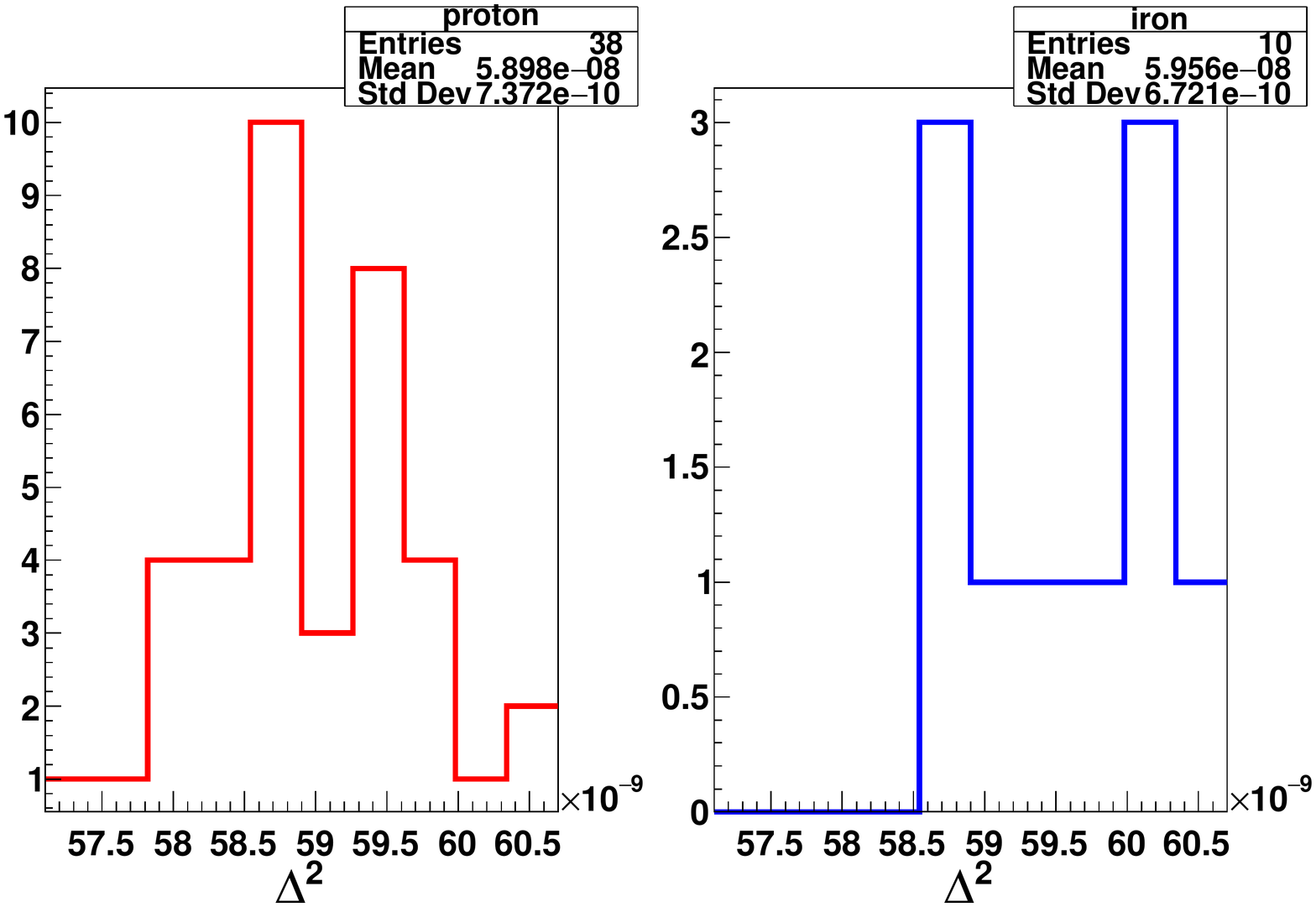}
    \caption{Same as Fig. \ref{fig:Recexample-nonoise}, but for an array with $D=750$ m and taking into account all detection uncertainties, including energy. (see text for details).}
\label{fig:Recexample}
\end{center}
\end{figure}

In Fig.\,\ref{fig:Recexample} we show the same input proton event as in Fig.\,\ref{fig:Recexample-nonoise}, but now in the sparser array and accounting for all detection uncertainties. One can see that the large separation between the distributions of $\Delta^2$ for proton and iron simulations in Fig.\,\ref{fig:Recexample-nonoise}, and which we attribute to the difference in the average missing energy of proton and iron events, has almost disappeared when accounting for the uncertainty in energy.

All simulations used in this work have a fixed azimuth angle of $\phi=90^\circ$, i.e. they all come from the North. We chose these showers since they have a larger probability of detection at the LOFAR site. We do not expect the azimuthal arrival direction of the shower to have a large impact on the effectiveness of the method to discriminate between different primaries. However, the amplitude of the electric field is experimentally known to depend on the azimuth angle of the showers \cite{LOFAR_expt,tunkarex}. For certain directions, for which the electric field amplitude is much smaller than others, noise would tend to have a larger impact on the detection. This azimuthal angle dependence is due to the geomagnetic contribution to the emission varying with $\sin\alpha$, where $\alpha$ is the angle between the shower axis and the geomagnetic field. Given the direction of the geomagnetic field at the LOFAR site, showers coming from the North ($\phi=90^\circ$) have larger fields ($\alpha$ closer to $90^\circ$) than showers coming from the South, leading to an observed \cite{LOFAR_expt} North-South asymmetry in the number of detected events at LOFAR, where a value of $\alpha=90^\circ$ (maximum geomagnetic contribution) would occur for a $\theta=67^\circ$ shower coming from the North. These showers produce a higher total electric field and have a larger SNR than those for instance from the South for the same energy. As we change the azimuth angle %$\alpha$ becomes smaller than $90^\circ$,
the net electric fields tend to diminish and, as a consequence, this slightly increases the influence of some experimental uncertainties on the discrimination, especially that due to noise.

%---------------------------------------------------------------------------------------
%---------------------------------------------------------------------------------------

%---------------------------------------------------------------------------------------
%---------------------------------------------------------------------------------------

%%%%%%%%%%%%%%%%%%%%%%%%%%%%%%%%%%%%%%%%%%%%%%%%%%%%%%%%%%%%%%%%%%%%%%%%%%%%%%%%%%%%%%%%%%%%%%%%%%%%%%%%%%%%%%%%%%%%%
\section{$X_{\rm max}$ reconstruction using information from radio detection}
\label{sec:xmaxuncertainty}

In this section we study the performance as a function of shower zenith angle of the most commonly used method to reconstruct \xmax based on information extracted from the radio detection of air showers. For this purpose we developed a variation of the method used in the LOFAR experiment \cite{lofarxmax}, where both radio and scintillator detector data are used for the reconstruction. In our simplified version only the radio signals of the event are used. Similar variations of this method have also been used to reconstruct \xmax with radio data collected at the AERA experiment, with encouraging results for showers with zenith angle $\theta\lesssim 60^\circ$ \cite{augerradiodetection-turin2017}. A comparison of the performance of our simplified method described below (labeled as method D) and others used in AERA can be found in \cite{augerxmaxmethods-ARENA2016}.

This method is also based on comparisons between the electric field measured in several antennas and that predicted in \zhaires Monte Carlo simulations having the same geometry and energy as the detected event, but with different primary compositions, spanning the whole $\xmax$ range. For this purpose we first calculate the quantity $\Sigma_s$ in Eq.\,(\ref{eq:sigma}), defined as the quadratic sum of the differences between the measured and predicted peak amplitude of the electric fields over all antennas with signal: 
\begin{equation}
\Sigma_s(f_s,\vec{r}_{\rm core})=\sum_{\rm antennas}
{\left[ \vert\vec{E}_{\rm data}\vert - f_s \cdot \vert\vec{E}_{{\rm MC}}\vert (x-x_{\rm core},\,y-y_{\rm core}) \right]^2}.
\label{eq:sigma}
\end{equation}  

Here $f_s$ is the energy scaling factor, $\vec{r}_{\rm core}=(x_{\rm core},\,y_{\rm core})$ is the position of the shower core in the simulation and $(x,\,y)$ is the position of each antenna. In contrast to $\Delta_s$ in Eq.\,(\ref{eq:newsigma}), here we only use the peak amplitudes $\vert\vec{E}_{\rm data}\vert$ and $\vert\vec{E}_{{\rm MC}}\vert$ for the measured and simulated electric fields, respectively.

%%%FROM DISCRIMINATION METHOD BEFORE SECTION INVERSION
As in the method described in Section \ref{sec:composition}, we firstly perform simulations of 50 proton and 50 iron-initiated showers, with the same energy and geometry of the event to be reconstructed.  Here we also have the option to take into account uncertainties in the energy and core position of the detected event by varying the values used for the position of the core ($\vec{r}_{\rm core}$) and the energy scaling factor ($f_s$). Only the minimum value of $\Sigma_s(f_s,\vec{r}_{\rm core})$ for each simulation, denoted simply as $\Sigma$, is used in the analysis. This corresponds to the values of $f_s$ and $(x_{\rm core},y_{\rm core})$ for which the simulated shower represents best the measured electric field. On the other hand, if one does not want to take detection uncertainties into account in the reconstruction, the fixed values $f_s=1$ and $\vec{r}_{\rm core}=0$ are used in all calculations. Since in this section we are interested in finding the minimum possible uncertainties in \xmax, i.e. those inherent to the methodology, this is the approach we used.

\begin{figure}[ht]
\centering
\includegraphics[width=0.9\textwidth]{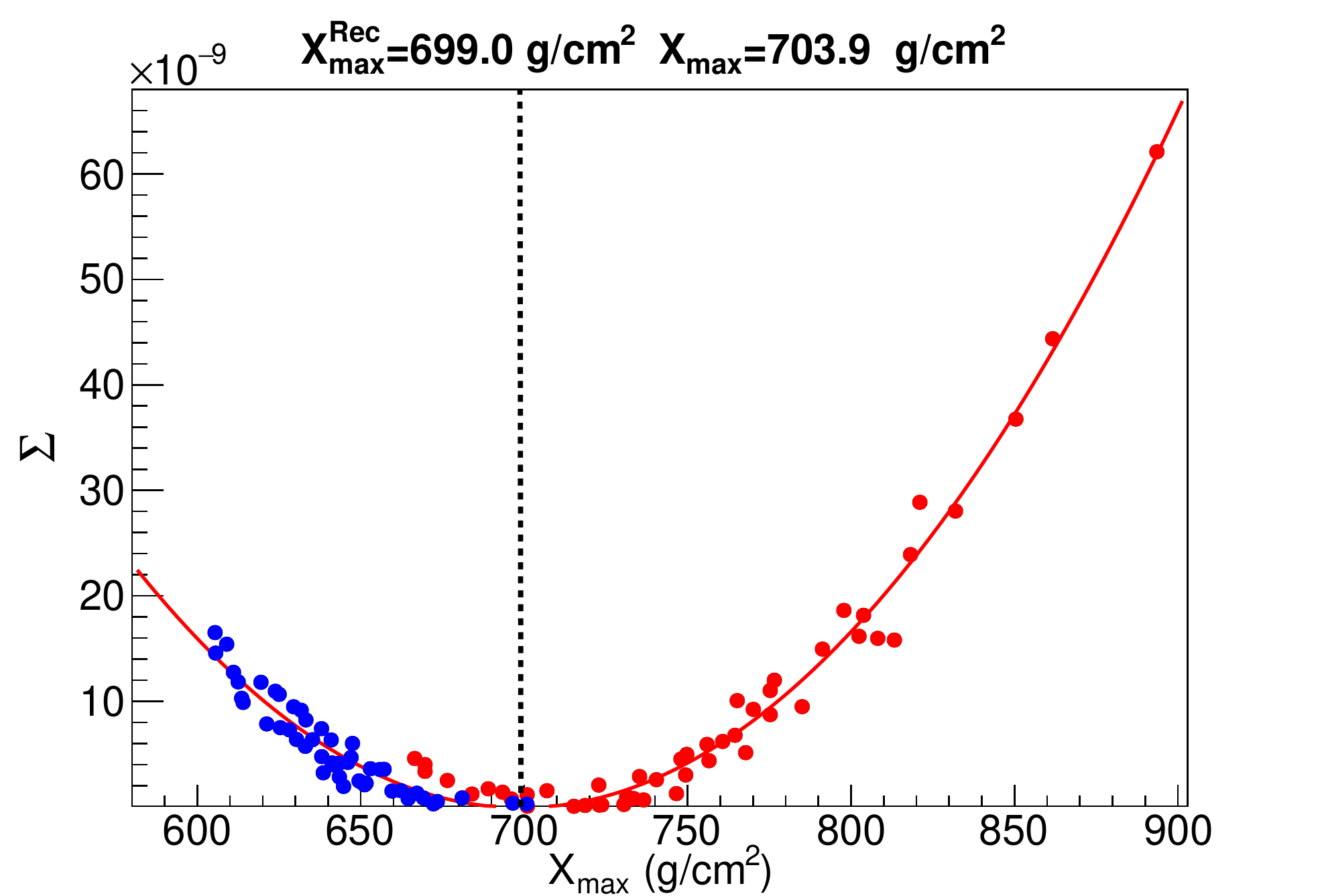}
\caption{Example of the procedure to reconstruct the depth of maximum of a $10^{18}$\,eV proton shower with $\theta=30\degree$. No detector uncertainties were folded into the simulated input event. The value of $\Sigma$ obtained in each of the 50 proton simulations (red dots) and each of the 50 iron simulations (blue dots) is plotted as a function of the $\xmax$ of each simulation. The red line represents a parabolic fit to the points to find the position of the minimum, which is the reconstructed $X^{\rm Rec}_{\rm max}$ of the event. The vertical dashed line represents the $\xmax=703.9\,{\rm g\,cm^{-2}}$ 
of the (simulated) input event. $X_{\rm max}^{\rm Rec}=699.0\,{\rm g\,cm^{-2}}$ denotes the value of the reconstructed $\xmax$ (see text for details on the reconstruction procedure).}
\label{fig:examplexmaxrec}
\end{figure}

To reconstruct the \xmax of the input event, the value of $\Sigma$ as a function of  $\xmax$ for each individual (proton and iron) simulation is plotted, and a parabolic fit is then performed. The position of the minimum of the parabola is taken as the reconstructed $\xmax$ of the event, denoted as $X_{\rm max}^{\rm Rec}$. In Fig.\,\ref{fig:examplexmaxrec} we show an example of the reconstruction procedure, where we used as input event to be reconstructed a (simulated) proton-induced shower with $E=10^{18}$\,eV and $\theta=30^\circ$. An ideal squared-grid array with distance between antennas $D=500$ m was used. We did not take into account any detection uncertainties and fixed both $f_s=1$ and $\vec{r}_{\rm core}=\vec{0}$ in Eq.\,(\ref{eq:sigma}). Each point in the plot corresponds to a single proton (red) or iron (blue) simulated shower of the 50 proton and 50 iron shower simulations used to reconstruct the input event. 

%-------------------------------------------------------------------------------------------------------
%-------------------------------------------------------------------------------------------------------

\begin{figure}[htb!]
\centering
\includegraphics[width=0.8\textwidth]{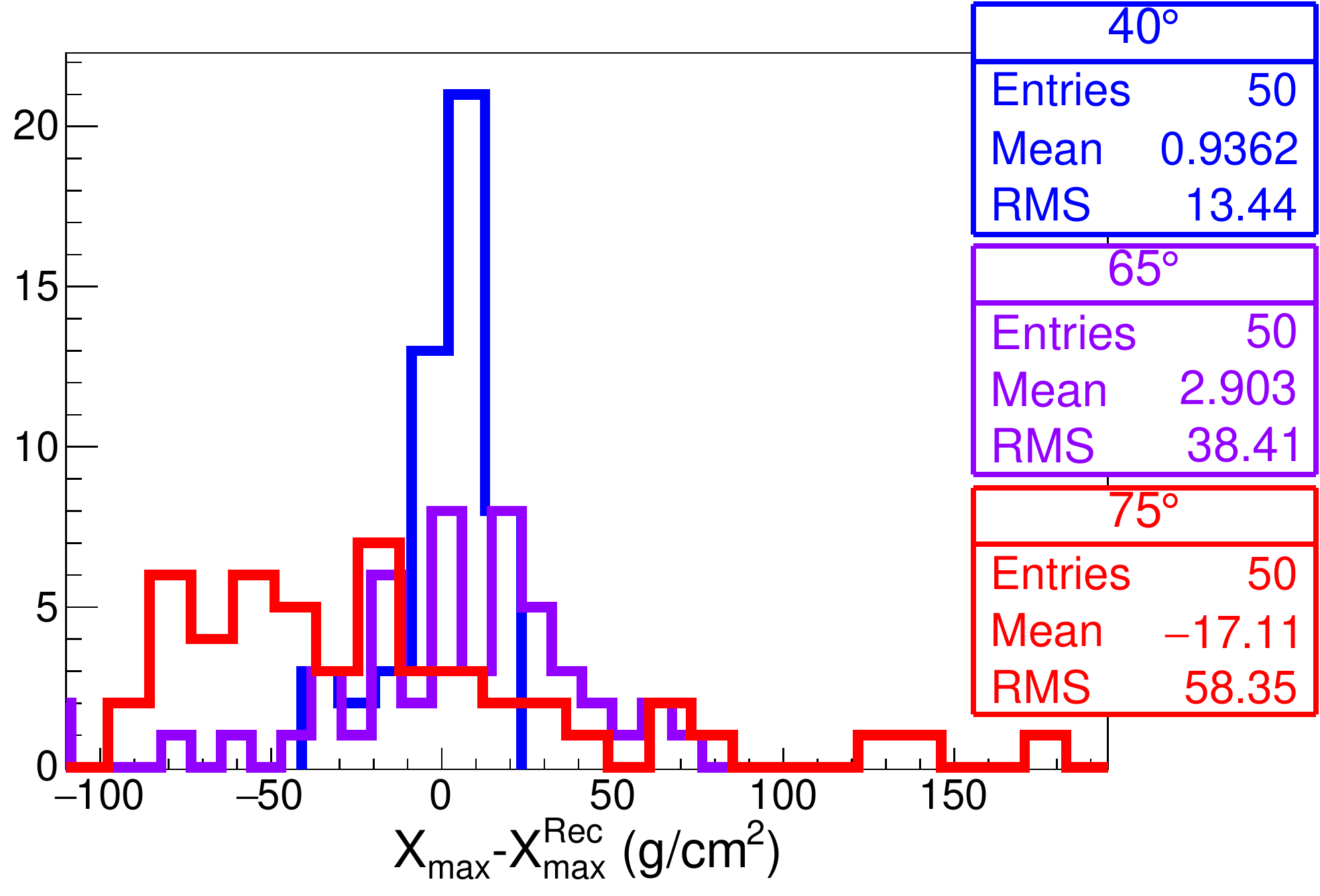}
\includegraphics[width=0.9\textwidth]{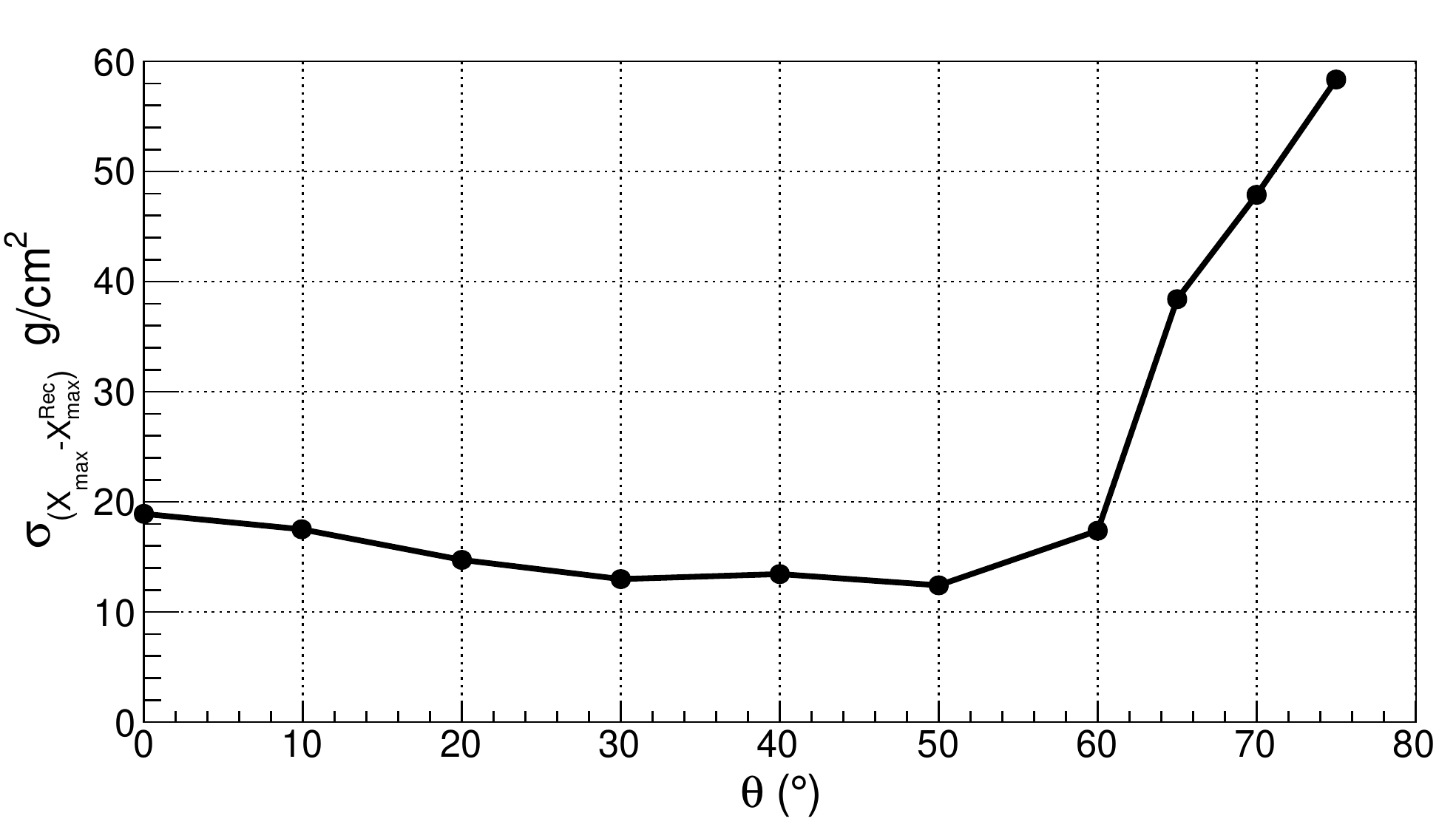}
\caption{Top: distributions of $\xmax-\xmaxrec$ for input showers with $\theta=40$, $65$ and $75\degree$ and $E=10^{18}$\,eV on an ideal squared-grid antenna array with $D=500$ m. Bottom: Uncertainty on $\xmaxrec$ (quantified as the RMS of the corresponding distribution of $\xmax-\xmaxrec$) as a function of shower zenith angle. No detection uncertainties were folded into the input events that were reconstructed.}
\label{fig:sigxmaxD500}
\end{figure}

Repeating the same procedure for showers at different zenith angles, we find that the quality of the $\xmax$ reconstruction, quantified using the difference between the $\xmax$ of the input event and the reconstructed $\xmaxrec$, depends strongly on the zenith angle of the event. This is shown in the bottom panel of Fig.\,\ref{fig:sigxmaxD500}, where we plot the RMS of the distributions of $\xmax-\xmaxrec$ when the $\xmax$ of 50 (simulated) input events are reconstructed for each value of $\theta$. The uncertainty on the reconstructed value of $\xmax$ is below $\sim 20\,{\rm g\,cm^{-2}}$ for $\theta\lesssim 60\degree$ rapidly increasing for more inclined showers and reaching $\sim 60\,{\rm g\,cm^{-2}}$ at $\theta\sim75\degree$. No detection effects or uncertainties were folded into the input events, and the plotted values of the RMS represent the minimum possible uncertainties, i.e. those inherent to the method. In the top panel of Fig.\,\ref{fig:sigxmaxD500} we also show three distributions of $\xmax-\xmaxrec$ for three different zenith angles.  

%-------------------------------------------------------------------------------------------------------
%-------------------------------------------------------------------------------------------------------

The loss of sensitivity to $\xmax$ as $\theta$ increases is mainly due to the combination of two effects that are responsible for the increase in uncertainty on $\xmaxrec$ shown in Fig.\,\ref{fig:sigxmaxD500}. 
The superposition of the Askaryan and geomagnetic contributions to the radio emission in air showers generates an asymmetric electric field pattern (footprint) that is sensitive to the shower longitudinal profile. For geometrical reasons, the size of the footprint on the ground is also sensitive to the distance between the bulk of the shower and the ground. Both observables, the size and the asymmetry of the footprint, are sensitive to $\xmax$ \cite{lofarxmax}. As the zenith angle increases, the shower develops higher in the atmosphere in a region of lower air density and this enhances the contribution of the geomagnetic emission mechanism with respect to that in less inclined showers (see Section\,\ref{sec:radioemission} and Refs.~\cite{scholten-driftvelocity,LOFAR_polarization}), while the Askaryan emission remains practically the same. This effect makes the ratio of geomagnetic to Askaryan emission larger the more inclined the shower is, and the field pattern on the ground becomes more symmetric around the shower core. As a consequence, for zenith angles $\theta>65^\circ$ the footprint is practically symmetric and the information on $\xmax$ contained in the asymmetric pattern is lost, making the reconstruction of $\xmax$ less constrained. Moreover, as the shower zenith angle increases, the size of the induced Cerenkov ring on the ground, where the signal is largest \cite{zhaires-uhf}, becomes less sensitive to $\xmax$, further decreasing the sensitivity of the pattern to $\xmax$. The reason for this is that showers produced higher in the atmosphere tend to have larger footprints on the ground, because the beamed radiation is projected on the ground from a larger distance. However, this is compensated by the fact that the Cerenkov angle is smaller due to the lower density of air at higher altitudes.

\begin{figure}[htb!]
\centering
\includegraphics[width=0.5\textwidth]{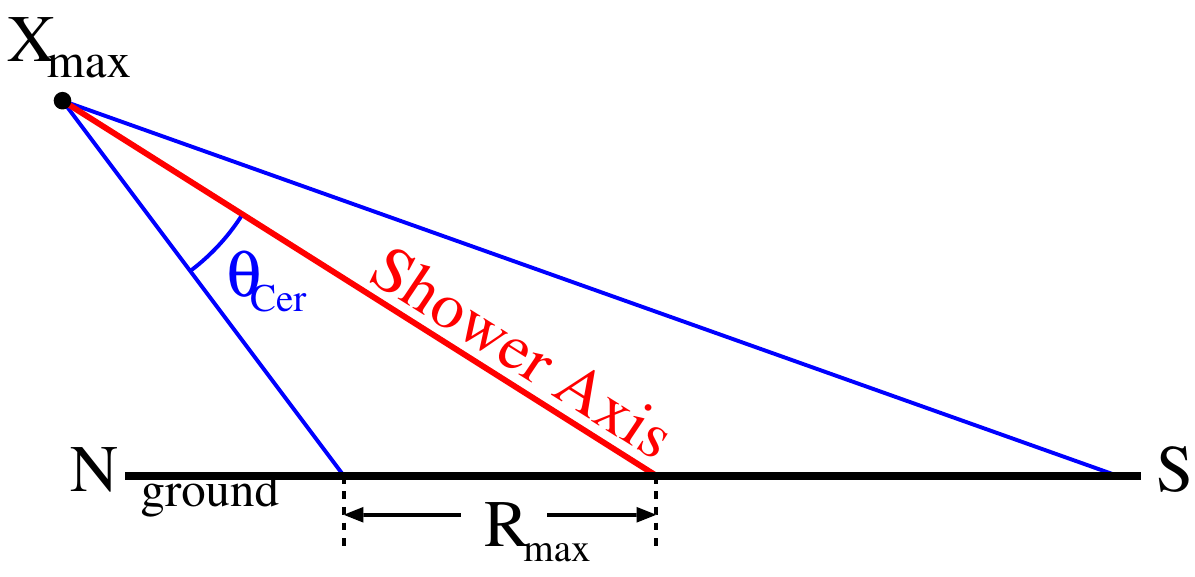}
\includegraphics[width=0.65\textwidth]{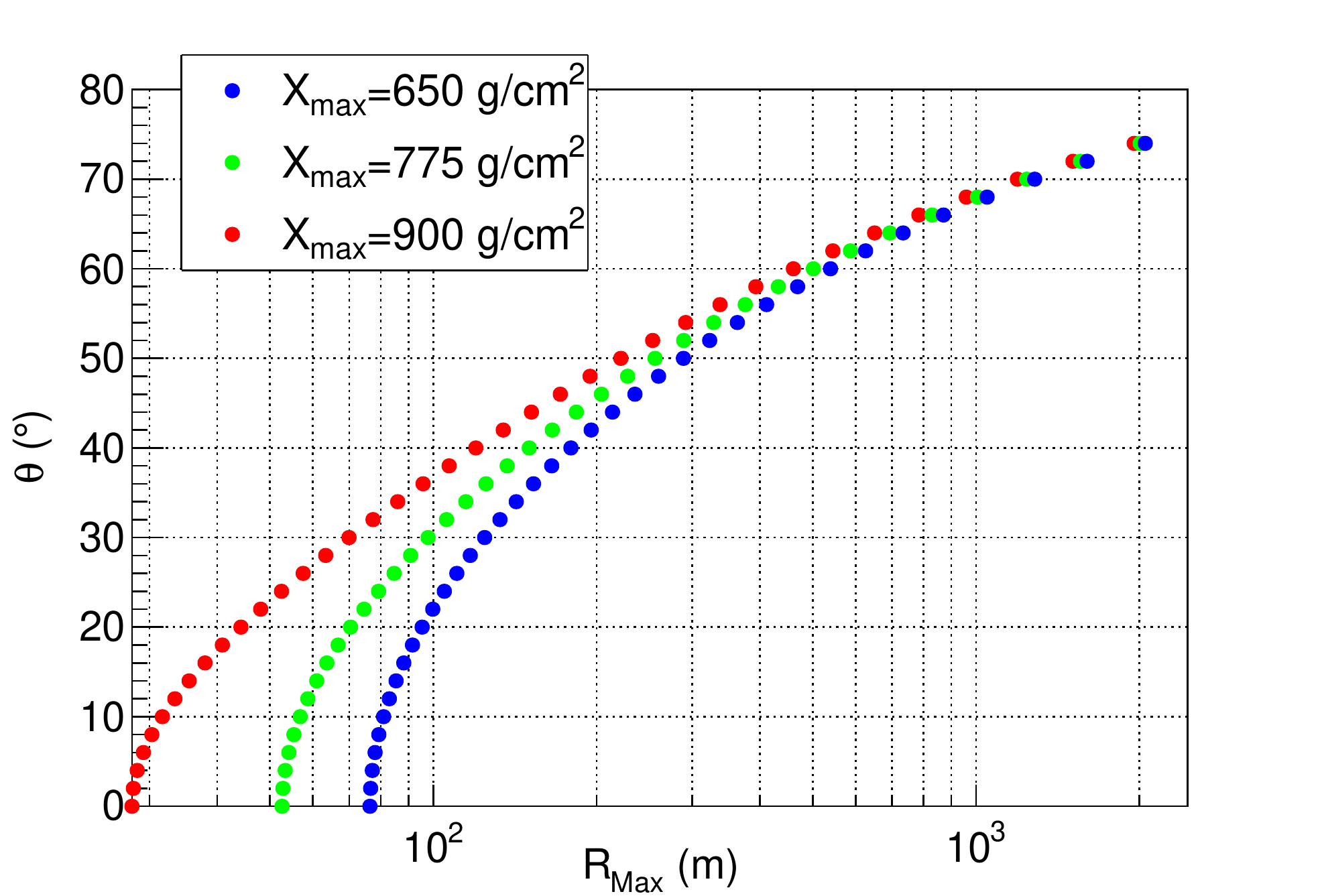}
\includegraphics[width=0.65\textwidth]{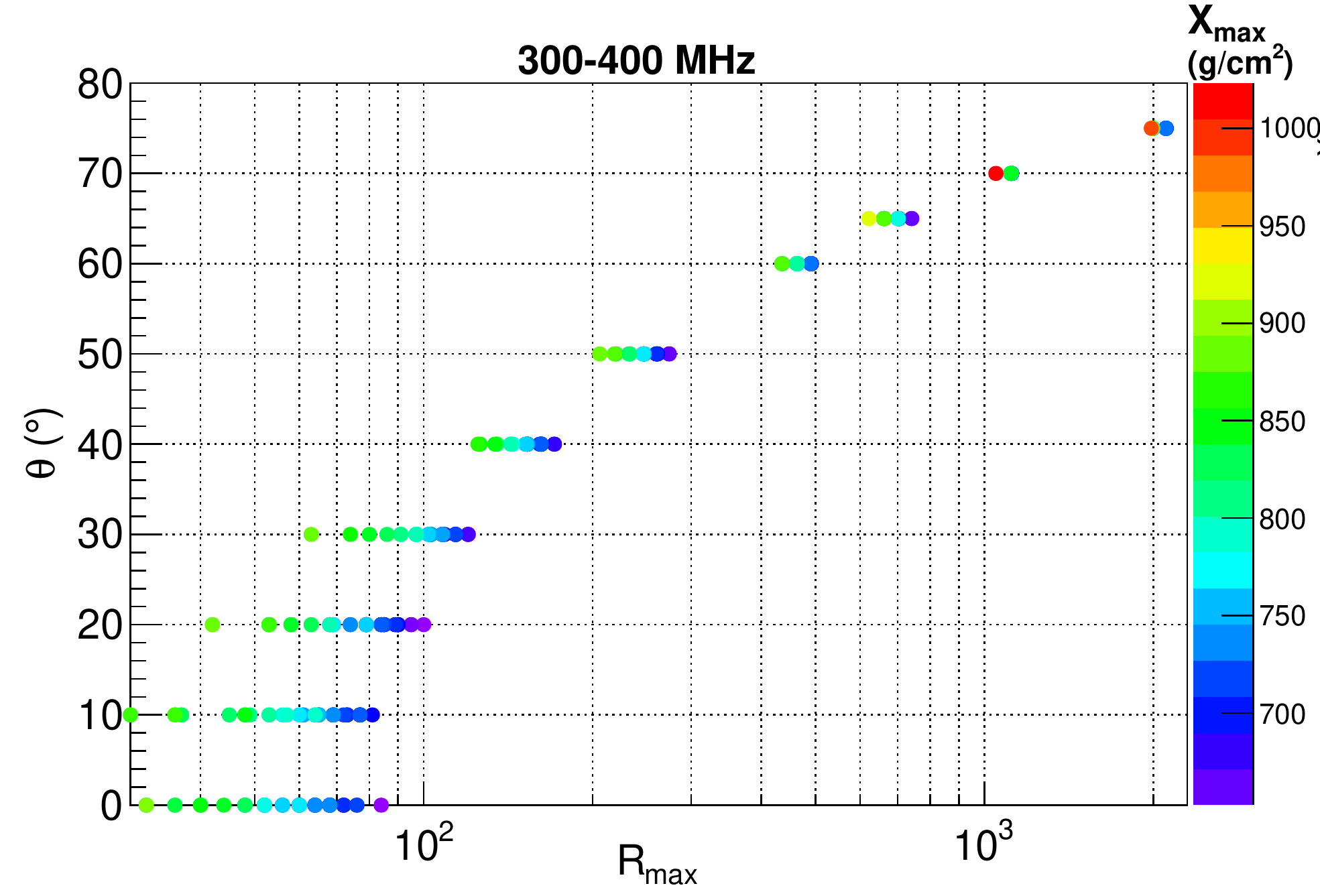}
\caption{Top: Sketch of the distance $\Rmax$ from the shower core to the Cerenkov ring, defined as the intersection of the cone of opening angle equal to the Cerenkov angle and vertex at $\xmax$. Middle: $\Rmax$ obtained analytically for several zenith angles and for 3 different values of $\xmax$, namely 650 (blue), 775 (green) and 900 ${\rm g/cm^2}$ (red), see text for more details. Bottom: Values of $\Rmax$ for different $\theta$ obtained from 50 full Monte Carlo simulations with \zhaires for each $\theta$. The color scale indicates the $\xmax$ of the simulated events.}
\label{fig:Rmax}
\end{figure}

The size of the footprint can be quantified in terms of $\Rmax$, that we define as the distance from the shower core to the Cherenkov ring along the same azimuth direction the shower comes from. In the case of the showers used in this work, all coming from the North, $\Rmax$ is the distance between the shower core and the point on the Cherenkov ring directly North of it. 
This ring is expected to appear at the intersection with ground of a Cerenkov cone centered at the position of \xmax  \cite{zhaires-uhf}. The axis of the Cerenkov cone is the shower axis, and its opening angle is the Cerenkov angle $\theta_{\rm Cher}$ at the $\xmax$ altitude. $\Rmax$ is depicted in the sketch in the top panel of Fig.\,\ref{fig:Rmax}. For a given $\xmax$ and zenith angle, $\Rmax$ can be obtained analytically with a model for the density and refractive index in the atmosphere, which allows one to calculate $\theta_{\rm Cher}$ at the $\xmax$ altitude. Values of $\Rmax$ calculated in this way, using the same refractive index model implemented in \zhaires \cite{zhaires-air}, are shown in the middle panel of Fig.\,\ref{fig:Rmax} for several zenith angles and for showers with 3 different fixed values of $\xmax$.
As expected from simple geometrical considerations, $\Rmax$ increases with $\theta$ as the distance from $\xmax$ to the ground increases. However, the relative difference between the values of $\Rmax$ obtained for the three different values of $\xmax$ decreases with $\theta$. Similar results were obtained in \cite{tesedaniel} using a different approach. This illustrates that $\Rmax$ becomes less sensitive to the distance to shower maximum as $\theta$ increases.  As a consequence, in the case of inclined showers, large variations on $\xmax$ lead to only small variations in the footprint size and $\Sigma$, making the determination of the minimum of the $\Sigma$ vs $\xmax$ curve, and thus \xmaxrec, very inaccurate. The values of $\Rmax$ can also be obtained directly from the Monte Carlo simulations. These are shown in the bottom panel of Fig.\,\ref{fig:Rmax}, where they are seen to follow the same trend as in the analytical calculation.

%---------------------------------------------------------------------------------------
%---------------------------------------------------------------------------------------

The one-dimensional shower model assuming that the bulk of radiation comes from a region around $\xmax$, although very simplistic, can describe well the position of the Cherenkov ring of showers with a distant $\xmax$, i.e. typically more inclined events. However, it represents a worse approximation for showers that have $\xmax$ closer to the ground, such as high-energy vertical showers. In this case, the one-dimensional shower approximation breaks down and the lateral spread of the shower becomes important \cite{zhaires-uhf}. In any case, the only purpose of the one-dimensional model was to show that the size of the radio footprint is less sensitive to Xmax in the case of inclined showers.

%%%%%%%%%%%%%%%%%%%%%%%%%%%%%%%%%%%%%%%%%%%%%%%%%%%%%%%%%%%%%%%%%%%%%%%%%%%%%%%%%%%%%%%%%%%%%%%%%%%%%%%%%%%%%%%%%%%%%
\section{Outlook and conclusions}
\label{sec:discussion}

We have shown that with the methodology presented in this work we can distinguish between light 
and heavy primaries on an event-by-event basis using information obtained from the radio detection 
of air showers. 

The interplay between the radio emission of each particle in the shower, coherence effects that depend on time, 
distance and frequency, time compression and beaming effects in the radio signal due to the variation of the 
refractive index with altitude along the line of sight, all paint a very complex picture. It is this same complexity
that generates the extra “degrees of freedom” in radio emission that are, at least in part, responsible for the ability 
to distinguish between light and heavy UHECR primaries. 

We have shown that reconstructions of $X_{\rm max}$ using radio detection exhibit larger uncertainties in the case 
of inclined events when compared to more vertical ones. This is due to the intrinsic characteristics of inclined showers, 
and makes $X_{\rm max}$-based composition studies of inclined events much more challenging. On the other hand, our methodology 
is more efficient at higher zenith angles $\theta\gtrsim 60^\circ$ and could be complementary to $X_{\rm max}$ composition analyses 
that have a good accuracy at lower zenith angles.

We have investigated the effect of detector uncertainties on the efficiency of our method. Even when a sparse array ($D = 750$ m) 
and upper limits to several experimental uncertainties (energy, core position and noise were included), $\sim 80\%$ of the events 
had their composition correctly inferred by the method for a zenith angle $\theta=65^\circ$. An efficient classification of 
primary cosmic rays into light or heavy on an event-by-event basis can help in the determination of the sources of UHECR, 
due to a better knowledge of the rigidity of the detected particles.

Our methodology, which currently uses only the measured peak electric field, can still be refined with maximum-likelihood, 
Bayesian analysis and iterative methods to take into account, not only the averages of the distribution of $\Delta$ in Eq.(1), 
but also its shape, in order to increase the discrimination efficiency, especially when large energy uncertainties 
are taken into account. 

In future works we will address the possibility of including the full-time pulse and not only the peak electric field in the method. 
This could increase the sensitivity to other parts of the shower development besides $X_{\rm max}$. We also intend to apply our method 
to real events in the future, as well as address what would be the optimal characteristics of future arrays of radio detectors for 
event-by-event composition determination based on our methodology.

%%%%%%%%%%%%%%%%%%%%%%%%%%%%%%%%%%%%%%%%%%%%%%%%%%%%%%%%%%%%%%%%%%%%%%%%%%%%%%%%%%%%%%%%%%%%%%%%%%%%%%%%%%%%%%%%%%%%%
\section{Acknowledgments}
%We would like to thank I.F.M. Albuquerque for her input and for proof-reading the manuscript.
W.R.C. is supported by Grants 2015/15735-1 and 2018/18876-3, São Paulo Research Foundation (FAPESP). J.A.-M. thanks Ministerio de Econom\'\i a (FPA 2015-70420-C2-1-R), Consolider-Ingenio 2010 CPAN Programme (CSD2007-00042), Xunta de Galicia (ED431C 2017/07), Feder Funds, 7th Framework Program (PIRSES-2009-GA-246806) RENATA Red Nacional Tem\'atica de Astropart\'\i culas (FPA2015-68783-REDT).
%%%%%%%%%%%%%%%%%%%%%%%%%%%%%%%%%%%%%%%%%%%%%%%%%%%%%%%%%%%%%%%%%%%%%%%%%%%%%%%%

%%%%%%%%%%%%%%%%%%%%%%%%%%%%%%%%%%%%%%%%%%%%%%%%%%%%%%%%%%%%%%%%%%%%%%%%%%%%%%%%

\end{document}